# A Room-Temperature Ferrotoroidic Material Exhibiting Magnetic Semiconductor Properties with Superhigh Hole Mobility

Jianyuan Qi, Shijie Xiong, Beining Ma, and Xinghai Shen*

*Fundamental Science on Radiochemistry and Radiation Chemistry Laboratory, Beijing National Laboratory for Molecular Sciences, Center for Applied Physics and Technology, College of Chemistry and Molecular Engineering, Peking University, Beijing 100871, P. R. China.*

The design and fabrication of room-temperature ferrotoroidic materials and magnetic semiconductors are recognized worldwide as a great challenge, and of both theoretical and practical importance in the field of condensed matter physics and information storage. Reported herein are ferrotoroidic crystal powder and film formed by supramolecular self-assembly based on uranyl and cyclodextrin, with the Curie temperature above room temperature. Experimental measurements and calculations demonstrate spontaneous vortex-like alignment of magnetic moments and furthermore a macroscopic long-range arrangement in the crystal, which breaks simultaneously space-inversion and time-reversal symmetries, exhibiting strong superexchange, spin-orbit coupling as well as anomalous Hall effect (AHE). The electrical measurements show the film with a superhigh carrier mobility of $(3.2 \pm 0.2) \times 10^3$ $cm^2V^{-1}s^{-1}$ and a Hall resistivity as high as 0.32 mΩ·cm at room temperature. This work is expected to pave greatly the applied research on new-generation magnetoresistive random access memory (MRAM), especially as flexible material.

## I. INTRODUCTION

A spontaneous vortex-like alignment of magnetic moments and furthermore a macroscopic long-range arrangement give rise to the so-called ferrotoroidicity, which breaks simultaneously space-inversion and time-reversal symmetries [1–7]. Ferrotoroidicity is regarded as the fourth primary ferroic order after ferromagnetism, ferroelectricity and ferroelasticity, and shows prosperous application in information storage [2–4]. Based on systematic theoretical research [8–11] and a few experimental studies [3–5,12], the main characteristics for a qualified ferrotoroidic material can be summarized as follows: (1) a vortex-like arrangement of atomic magnetic moment to from toroidal moment; (2) long-range ferrotoroidic order; (3) simultaneous breaking of space-inversion and time-reversal symmetries; (4) strong spin-orbit coupling; (5) magnetoelectric effect. So far, various inorganic materials have been explored to demonstrate their ferrotoroidicity. The ferrotoroidic domains of $LiCoPO_4$ [1] and $NdB_4$ [5] were found to prove their ferrotoroidicity. Other candidate materials like $Ba_6Cr_2S_{10}$ [4] and $MoX_3$ [13] are regarded as one-dimensional antiferromagnetic ferrotoroidic materials. It seems difficult to obtain a qualified ferrotoroidic material. Especially, the room-temperature ferrotoroidicity, which relies on strong spin-orbit coupling and superexchange to enhance magnetic interactions, is still highly expected for the purpose of practical application.

Very recently, our research group prepared a $UO_2^{2+}$- γ-cyclodextrin (γ-CD) supramolecular crystal (**U2** crystal) by self-assembly method [14]. After photoreduction or irradiation reduction, most $UO_2^{2+}$ is reduced to $UO_2^{+}$, which can remain stable in the **U1@U2** crystal. The room-temperature magnetism and a band gap of 1.91 eV suggest that the **U1@U2** crystal should be a room-temperature magnetic semiconductor. It is particularly noteworthy that the magnetic $UO_2^{+}$ ions are arranged in a vortex-like pattern between cyclodextrin units which are connected into tubular structure [14]. Therefore, one can infer that the **U1@U2** crystal might be a ferrotoroidic material due to its typical structural characteristics. If so, it is necessary to prove its ferrotoroidicity through the experimental characterization and theoretical calculation referring to the aforesaid main characteristics. Furthermore, it is of great importance to investigate some properties, *e.g.* magnetic semiconductivity, carrier mobility and anomalous Hall effect (AHE) at room-temperature, of the **U1@U2** crystal relating to the application in information storage. To enhance the application potential in electric device fabrication, it is necessary to explore the preparation on thin films [15,16]. This is equally applicable to the **U2** and **U1@U2** films.

Magnetic semiconductors have attracted much attention recently. They are expected to be applied extensively in spintronics since both charge and spin degrees of freedom can be manipulated by external stimuli [17,18]. In 2005, *Science* published 125 most challenging scientific problems, one of which is whether magnetic semiconductors can be produced at room-temperature [17]. In the past two decades, researchers have been pursuing this elusive goal on preparing room-temperature magnetic semiconductor materials. One strategy is on designing diluted magnetic semiconductors (DMSs), in which the magnetism is introduced through magnetic dopants [18,19]. Despite great efforts, the goal of realizing all key components within bulk semiconductors remains unattained thus far [17]. Another promising solution should rely on the intrinsic ferromagnetic semiconductors with high $T_C$, in which magnetic ions are periodically arranged [20,21]. Notably, the supramolecular self-assembly is a good method to construct semiconductors with long-range ordered ferroic nature [22,23]. For example, a ribbon-shaped ferroelectric assembly [22] and a ferroelectric-antiferromagnetic semiconductor [23] are separately prepared by a supramolecular assembly method. To the best of our knowledge, supramolecular self-assembled room-temperature ferromagnetic or ferrimagnetic semiconductors have not been reported yet [23].

Semiconductors are milestones in modern electronic devices, including bipolar transistors, complementary metal oxide semiconductor (CMOS) and so on [24]. However, many of the potential electronic applications are limited by the lack of p-type semiconductors with high carrier mobility [25–27]. Currently, the mobility of p-type semiconductors lags far behind that of n-type semiconductors. The reported hole mobility values are only 120 $cm^2V^{-1}s^{-1}$ for SiC [28], 800 $cm^2V^{-1}s^{-1}$ for InSe [29], and 1350 $cm^2V^{-1}s^{-1}$ for black phosphorene [30] at room temperature. Therefore, high-mobility p-type semiconductors matching with n-type semiconductors are extremely anticipating.

The first objective of this work is to confirm the ferrotoroidicity of the **U1@U2** crystal powder and film, and elucidate the ferrotoroidic mechanism. The second purpose is to explore the application prospect of the ferrotoroidic material by examining whether it is with strong magnetic semiconductivity, superhigh hole mobility and AHE phenomena at room temperature.

## II. MAIN CONTENT

### A. The preparation and structural characterization on **U2** and **U1@U2** films

To fabricate the film based on uranyl and γ-CD, it is necessary to construct favorable growth environment for supramolecular self-assembly. Herein, we report on an optimized method to prepare $UO_2^{2+}$-γ-CD film (**U2** film) on Si substrate by spin-coating [25]. The synthesis details are described in Supporting Information (SI). In the process of spin-coating, γ-CD can coordinate with metal ions and undergo supramolecular self-assembly to form crystals, which are connected and stacked to form a polycrystal film (Fig. 1a). The thickness of a **U2** film can be tuned by adjusting the spin coating speed and precursor solution concentration (see the experimental section in SI). Specifically, the film thickness decreases on decreasing the precursor solution concentration and increasing the spin coating speed. For a precursor solution at 75 mM, the film thickness is reduced from 1.4 to 0.8 μm by increasing the spin-coating speed from 2000 to 8000 rpm (Fig. S1). According to the crystal powder X-ray diffraction (PXRD) data, the diffraction peaks of both the **U2** film (Fig. S2a) and **U2** crystal powder are in accordance with the theoretical spectrum [14], which proves that the structure of the crystal in the **U2** film keeps the same as that in the **U2** crystal powder. The crystal belongs to tetragonal group and exhibits a sandwich-type coordination structure, where $UO_2^{2+}$ and $Cs^+$ ions are alternately arranged between cyclodextrin molecules to form a sixteen-membered heterometallic ring [14]. More details of the **U2** crystal can refer to the published work [14]. Scan electron micrograph (SEM) photos show that the **U2** film is actually composed of numerous small crystals, which are interlaced and stacked with each other (Fig. 1b and 1c). The good uniformity can be confirmed by transmission electron micrograph (TEM) photos (Fig. 1d and 1e). High-resolution TEM reveals the main lattice fringe of the **U2** film is 1.201 nm, assignable to the (111) crystal plane (Fig. 1e). One finds that the **U2** film retains the basic crystal structure of **U2** crystal powder, despite the different dominant crystal planes [14].

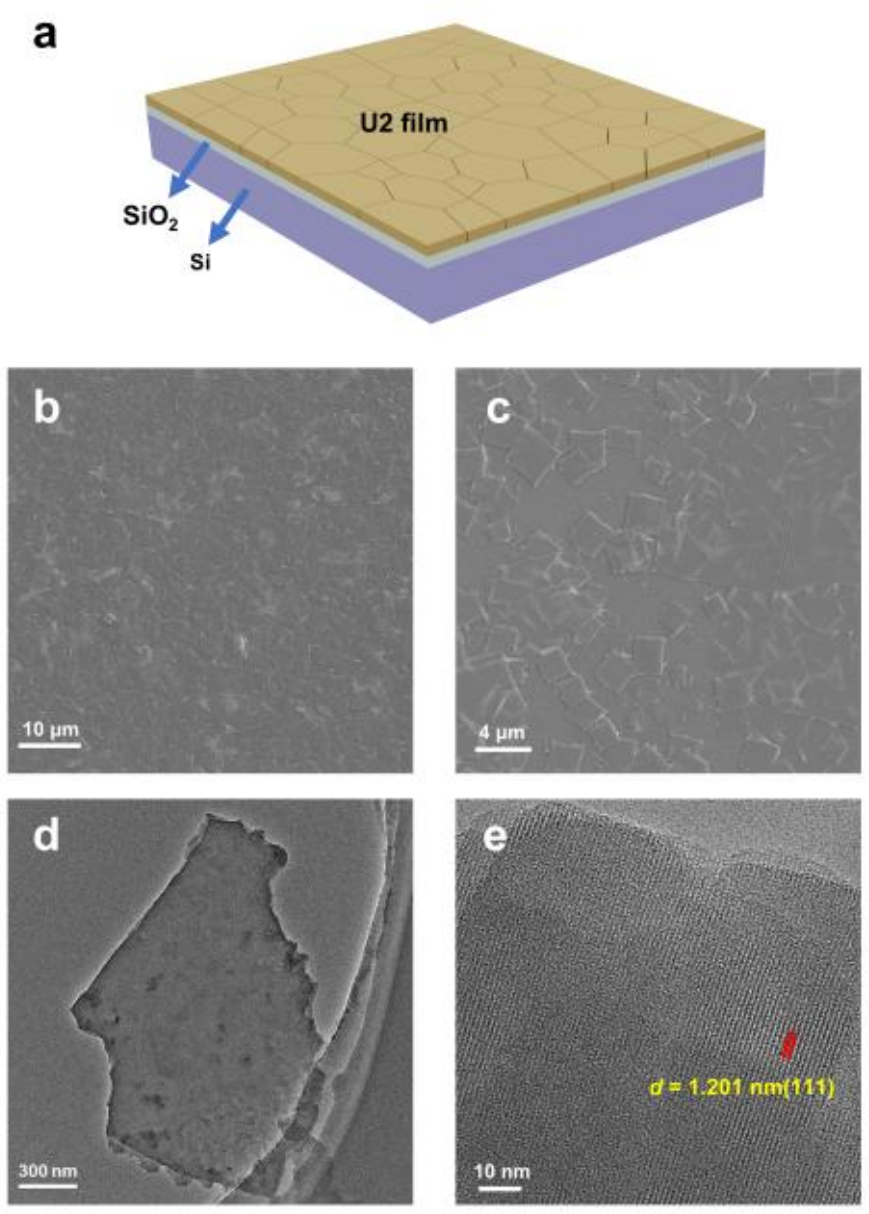

FIG. 1. The structure of the **U2** film. (a) Schematic diagram on the Si substrate. (b) Scanning electron micrograph (SEM) of the center area. (c) SEM of the edge area. (d) Low-resolution transmission electron micrograph (LRTEM). (e) High-resolution transmission electron micrograph (HRTEM).

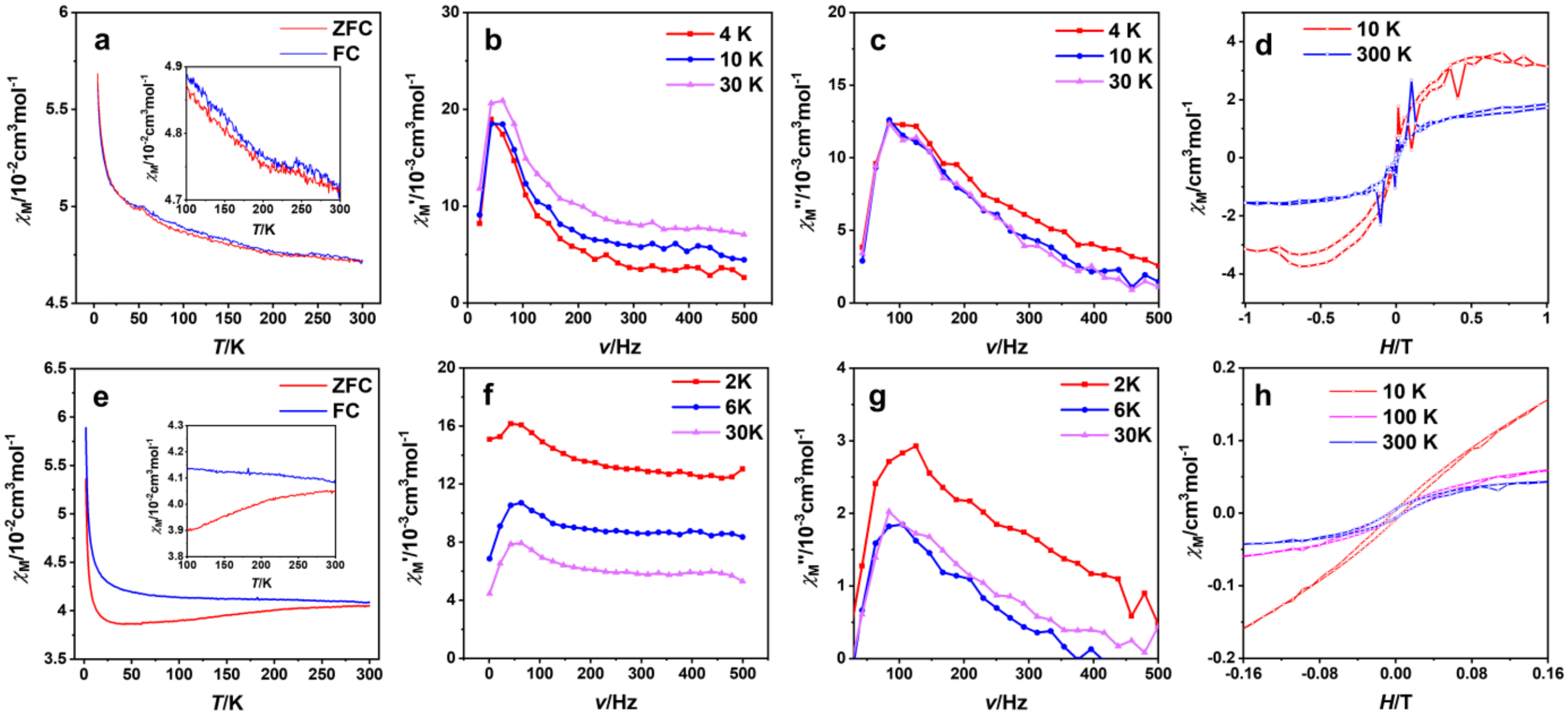

FIG. 2. Magnetic properties of the **U1@U2** film and crystal powder. (a) ZFC and FC curves of the **U1@U2** film. (b) Real and (c) imaginary parts of AC susceptibility of the **U1@U2** film at different temperatures. (d) *M-H* curves of the **U1@U2** film. (e) ZFC and FC curves of the **U1@U2** crystal powder. (f) Real and (g) imaginary parts of AC susceptibility of the **U1@U2** crystal powder at different temperatures. (h) *M-H* curves of the **U1@U2** crystal powder.

Similar to the reduction strategy for the **U2** crystal powder in previous work [14], the irradiation reduction was applied to the **U2** film resulting in the coexisting $UO_2^+$ and $UO_2^{2+}$ in the **U1@U2** film. For the irradiation method, the film received an absorbed dose of 600 kGy. After reduction, the **U1@U2** film maintains the good crystallinity and uniformity according to the energy dispersive spectroscopies (EDSs) and the TEM photos (Fig. S3). One can estimate from the X-ray photoelectron spectroscopy (XPS) data that 69.2% of the hexavalent uranyl in the film becomes the pentavalent (Fig. S4). The magnetism can be verified by electron paramagnetic resonance (EPR) spectra (Fig. S5). There are no signals in the EPR spectrum before reduction, while characteristic signal peaks do appear at $g$ = 2.015 and 2.006 after irradiation reduction (Fig. S5). This indicates that the **U1@U2** film is magnetic at room temperature.

The structure of the **U1@U2** film keeps stable after most $UO_2^{2+}$ ions are reduced to $UO_2^+$. This is confirmed by the powder X-ray diffraction (PXRD) spectra (Fig. S2), in which the peak positions and intensities remain almost unchanged after reduction, even if exposure of the film in air for a long time. More importantly, the crystallinity of the **U1@U2** film is improved compared with that of the **U1@U2** crystal powder, as can be seen from the strong and sharp peaks of PXRD for the former. Therefore, an efficient approach preparing large-scale flat **U1@U2** films with room-temperature magnetism is established successfully.

## B. Time-reversal and space-inversion symmetries breaking in U1@U2 film

The primary ferrotoroidic candidate materials reported to date, along with their key characteristics, are summarized in Tab. S1. One of the most important evidences confirming their ferrotoroidicity is the simultaneous breaking of time-reversal and space-reversion symmetries, which can be proved by magnetic characterization and second-harmonic generation (SHG) spectroscopy, respectively. It was found from the SHG spectra that the **U1@U2** crystal powder breaks the space-inversion symmetry in our previous work [14]. In this work, it is further demonstrated that both the **U1@U2** film and crystal powder break simultaneously the time-reversal and space-inversion symmetries. In addition to EPR, more magnetic properties of the **U1@U2** film can be demonstrated by magnetic transport measurements using magnetic property measurement system (MPMS). The curves of zero-field-cooling (ZFC) and field-cooling (FC) at 1000 Oe are not intersected within wide range of temperatures from 4 to 300 K (Fig. 2a), meaning that the thermal energy is incapable of disturbing the magnetically ordered state [31–33]. The Curie temperature ($T_C$) beyond 300 K suggests strong coupling among magnetic ions in the crystal. Both the real ($\chi'$) and imaginary ($\chi''$) parts of alternating current (AC) magnetic susceptibility exhibit frequency-independent peaks, which are not influenced by temperature, demonstrating a strong magnetic order in the **U1@U2** film (Fig. 2b and 2c). The **U1@U2** film presents hysteresis loops in *M-H* curves at 10 K and 300 K (Fig. 2d), indicating the transition temperature is above 300 K. This is in good accordance with ZFC-FC curves and AC measurements. It is worth mentioning that due to the extremely small mass of the film, the measured magnetic properties are relatively weak. In order to better characterize the magnetic properties of the crystal, the magnetic measurement of the **U1@U2** crystal powder by irradiation reduction is also carried out (Fig. 2e-h). The measured magnetic signal of crystal powder is stronger and smoother because of a greater mass. Although the magnetic properties of **U1@U2** film are weaker than those of the crystal powder, they agree with each other very well. From Fig. 2h, it can be found that there is a clear hysteresis loop at 300 K with a coercivity ($H_c$) of 80 Oe. For the **U1@U2** crystal powder by photoreduction, the magnetic spectra are also recorded (Fig. S6), which shows strong magnetic properties similar to those by irradiation reduction (Fig. 2e-h).

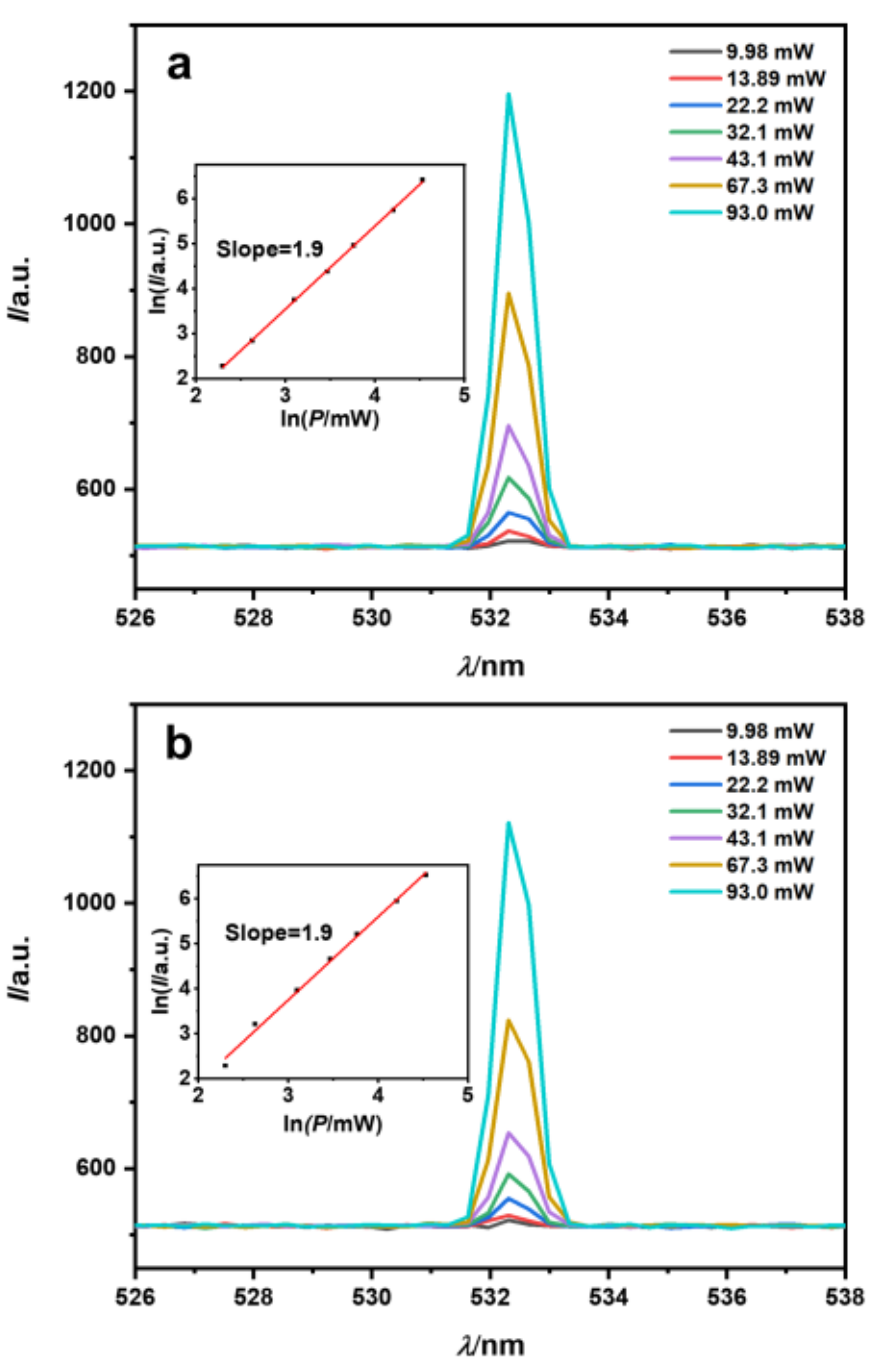

FIG. 3. Power-dependent SHG spectra of the **U2** (a) and **U1@U2** (b) films.

Fig. 3a exhibits the power-dependent SHG spectra of the **U2** film, where the frequency-doubling peaks at 532 nm (excitation wavelength: 1064 nm) are observed. As the power increases, the intensity of the SHG signal increases exponentially. The inset in Fig. 3a shows a logarithmic plot of the intensity versus the power, and the straight line with a slope of 1.9, close to the theoretical value of 2.0, confirming a real SHG signal [34]. Combined with the chirality nature of cyclodextrin [35], it can be concluded that the **U2** film breaks the space-inversion symmetry. After reduction, the **U1@U2** film retaining the crystal structure, shows power-dependent SHG spectra and a straight line of logarithmic plot of the intensity against the power similar to those before reduction (Fig. 3). It is worth mentioning that the intensity of the SHG signal of the **U1@U2** film is 1~2 orders of magnitude higher than that of the **U1@U2** crystal powder [14], which indicates the film has a stronger ordered orientation structure and a lower scattering loss [36].

C. Density function theory (DFT) calculations

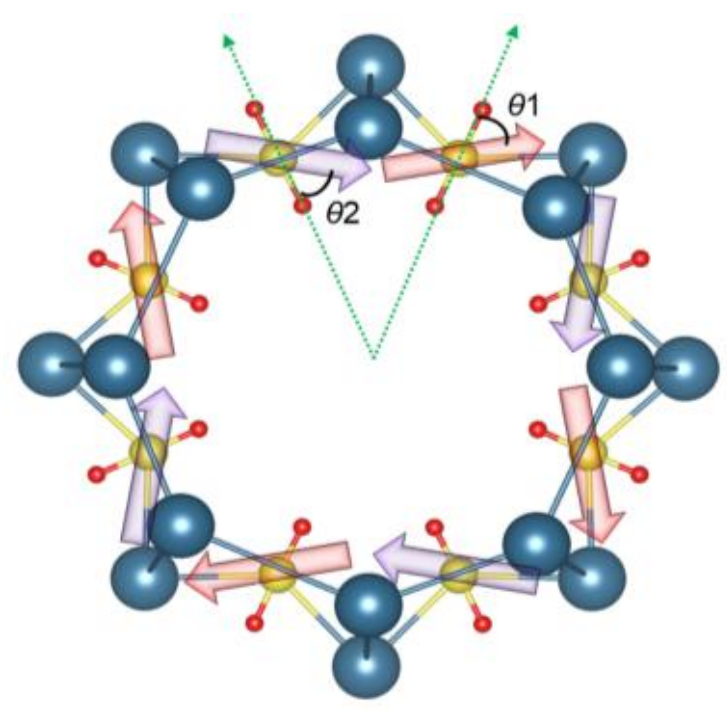

FIG. 4. Magnetic coupling model in a **U1** microstructural unit, indicating the formation of toroidal moment. *θ*1 and *θ*2 are the angles between magnetization direction and O=U=O direction.

The ferrotoroidicity with $T_C$ above room temperature has not been reported [17,37]. To analyze the reason for the high Curie temperature of the **U1@U2** crystal, the crystal model **U1** (all $UO_2^{2+}$ in **U2** powder crystal would be reduced to $UO_2^+$) is simulated by DFT calculations (Tab. S2). According to the calculation results, the spin-orbital coupling constant $\zeta$ is 2164.5 $cm^{-1}$, suggesting strong coupling of spin and orbital magnetic moments of $UO_2^+$. Combining strong spin-orbital coupling, the total magnetic moment of $UO_2^+$ is orientated in real space. The 5*f*-5*f* superexchange coefficient ($J$) between adjacent $UO_2^+$ ions reaches as high as 7.8 $cm^{-1}$, greater than that of $Np^V$=O−$Np^{VI}$ in $\{Np^{VI}O_2Cl_2\}\{Np^VO_2Cl(thf)_3\}$ [38]. The high $J$ value indicates a strong magnetic coupling of $UO_2^+$ ions in crystal [39], which is large enough to keep $T_C$ beyond 300 K. From superexchange calculations, the direction of vector sum of adjacent magnetic moments is along the line linking of neighboring U atoms, suggesting the head-to-tail magnetic moments coupling model, which refers to the toroidal moments (Fig. 4). The strong superexchange in a noncentrosymmetric crystal indicates the intrinsic magnetoelectric coupling effect [2,40].

D. Mechanism of room-temperature ferrotoroidicity

The ferrotoroidicity in this work arises from the formation of supramolecular self-assembled crystals (Fig. 5a). The supramolecular self-assembly of chiral cyclodextrins [35] with uranyl ions forms crystals with spatial-inversion symmetry breaking. Subsequently, two precise reduction methods as mentioned above are used to reduce $UO_2^{2+}$ to $UO_2^{+}$. The magnetism is thus introduced making the time-reversal symmetry broken. On the basis of in-depth analysis, one can propose a five-step coupling mechanism from magnetic moments to forming finally a ferrotoroidic material (Fig. 5b). In the first step, spin-orbital coupling provides uranium atoms a total angular moment presenting a coupled magnetic moment, which can be orientated in 3D real space through the special tetra-coordination in $\gamma$-CD. The second step is the formation of toroidal moment $\vec{T}$ caused by the vortex alignment of coupled magnetic moments and the superexchange. The third step concerns the coupling of toroidic moments $\vec{T}$ along the 1D tube and therefore the formation of the long-range ferrotoroidic moment $\sum\vec{T}$. The fourth step is focused on the further coupling of long-range ferrotoroidic moment $\sum\vec{T}$, by which ferrotoroidic domains are formed. Due to the chirality of $\gamma$-CD, the magnetic moments of electrons located on neighboring 1D tubes are with different directions, indicating opposite signs of neighboring long-range ferrotoroidic moments $\sum\vec{T}$. In the fifth step, different ferrotoroidic domains are further correlated together and the ferrotoroidic material is finally formed.

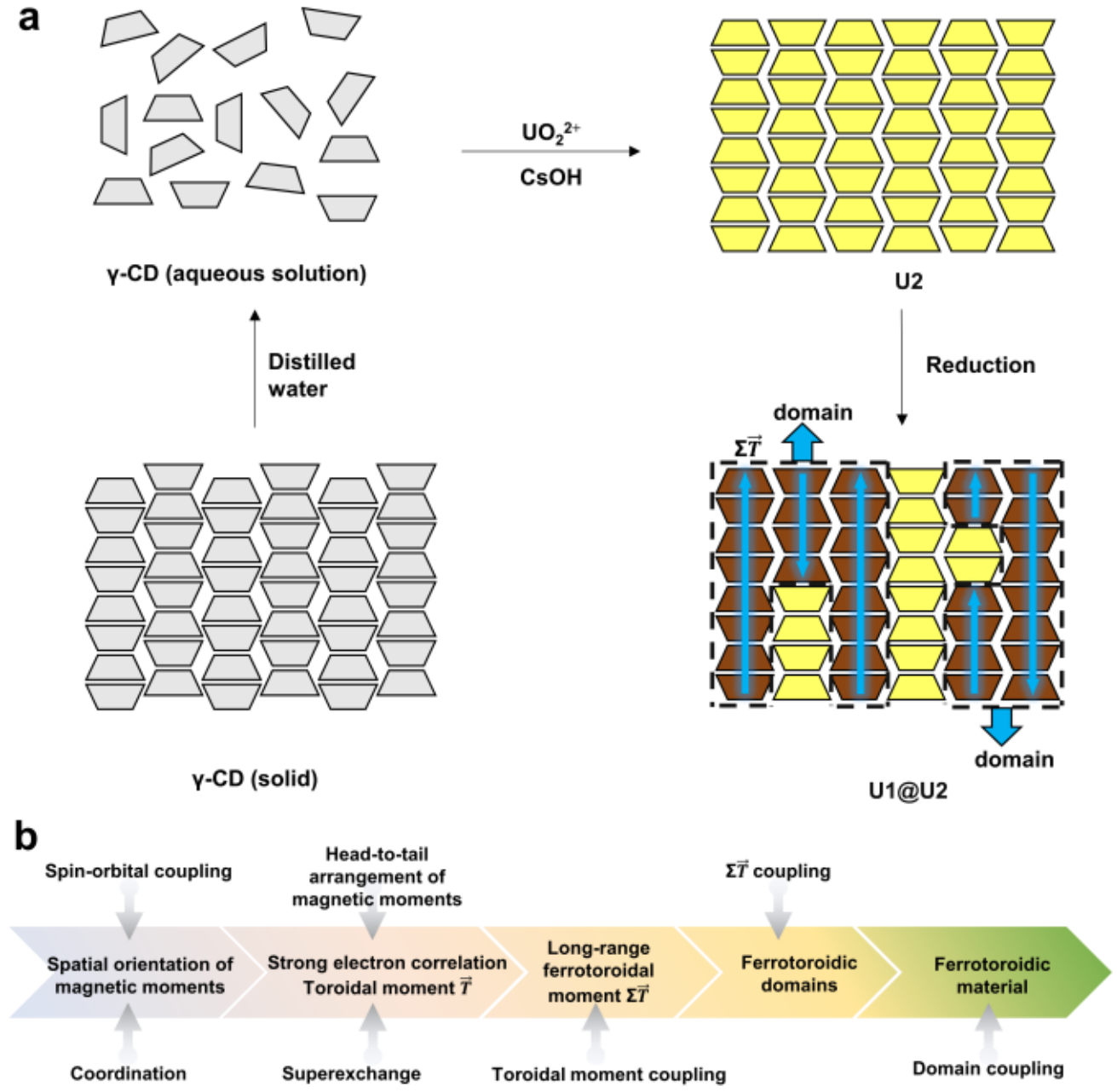

FIG. 5. Schematic diagram of the formation process and mechanism of ferrotoroidicity. (a) The formation process of ferrotoroidic domains. The crystal of natural γ-CD has dislocation among neighboring tubes while **U2** does not have this dislocation. For clarity, $UO_2^{2+}$, $UO_2^{+}$ and $Cs^{+}$ are omitted. (b) Five-step mechanism of ferrotoroidicity formation. The step-by-step coupling interactions lead to correlation structures from microscopic to macroscopic levels.

## E. Semiconductor properties of **U1@U2** film

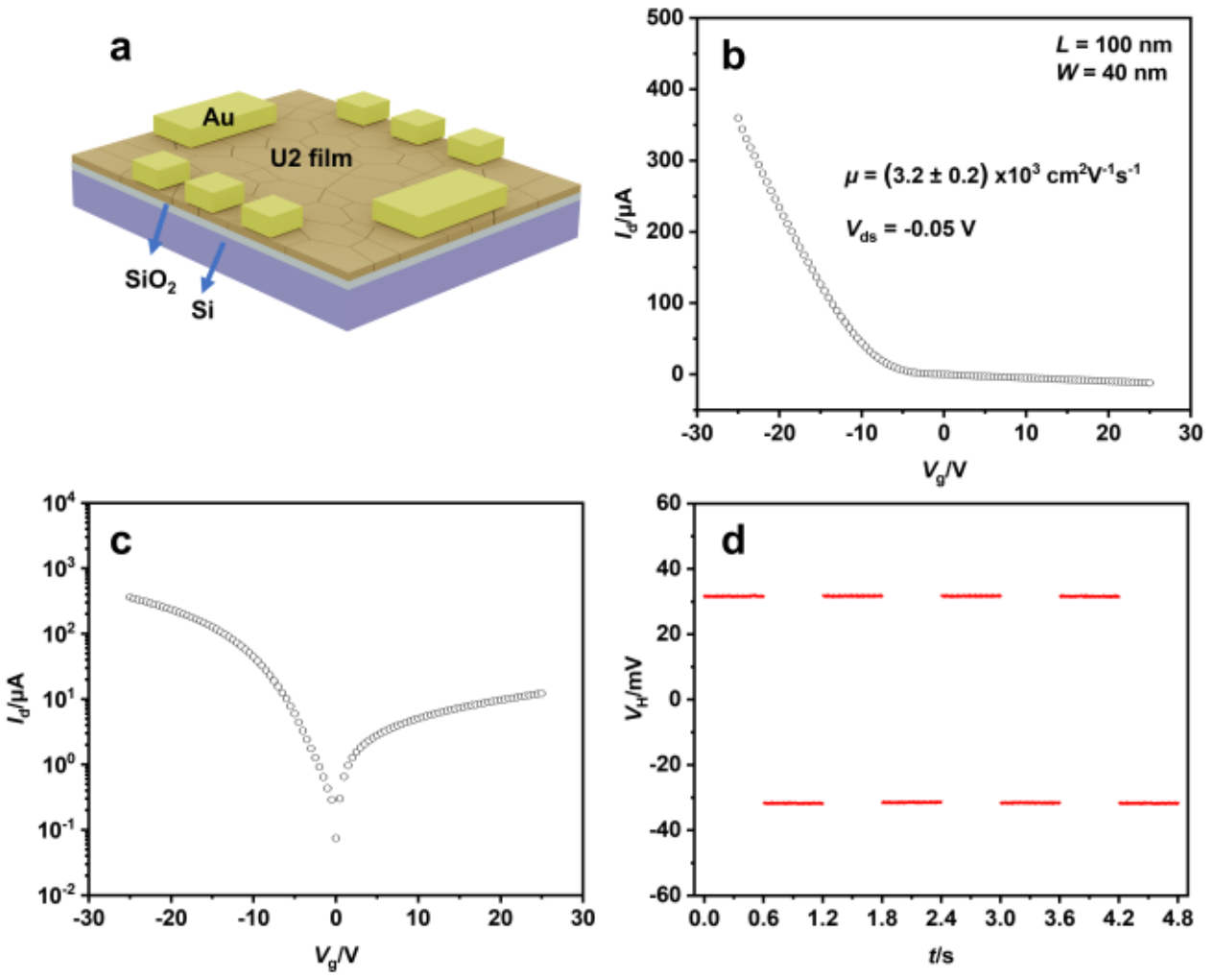

FIG. 6. The electrical characterization of the Hall device based on the **U1@U2** film. (a) The schematic diagram of the Hall device on Si substrate. (b) The transistor transfer curve; (c) Logarithmic treatment of the transistor transfer characteristic. (d) The AHE phenomena. The current is set to 10 mA and then to -10 mA, which is cycled four times with 0.6 s of intervals.

In order to verify the semiconductor properties, the Hall devices based on **U1@U2** films are fabricated through the micro-nano processing technology (Fig. S7), and the characteristic curves of field-effect transistors (FETs) are measured. Fig. 6a illustrates the schematic diagram of the **U1@U2** film-based FET, which exhibits a bottom-gate, top-electrode structure. The measured thickness of the $SiO_2$ layer is 4.73 ± 0.03 nm by ellipsometry (Fig. S8). The SEM image of the Hall electrode morphology is measured (Fig. S9), revealing a clear, high-resolution Hall bar pattern with 100 μm in length and 40 μm in width. The aspect ratio of the Hall bar is 2.5. The transfer characteristic curve of FET is displayed in Fig. 6b, which shows conduction onset at negative gate voltages and cutoff at positive gate voltages, corresponding to the characteristics of p-type semiconductor-based transistors. Moreover, the carrier mobility of the device is measured to be $(3.2 \pm 0.2) \times 10^3$ $cm^2V^{-1}s^{-1}$ ($V_{ds}$ = -0.05V) according to the transfer characteristic curve. To the best of our knowledge, the carrier mobility of the **U1@U2** film exceeds dramatically those of existing p-type semiconductor materials [41–43], which indicates a great potential in the application on electronic devices. The superhigh hole mobility can be attributed to the long-range ordered hole transport channel constructed by supramolecular assembly, which reduces lattice scattering. Moreover, there exists strong spin-orbit coupling in uranium atoms, which induces valence splitting and effective mass reduction, thereby promoting the movement rate of holes [44]. Fig. 6c further processes $I_d$ logarithmically to amplify the subthreshold region of the device. The linear region corresponds to the subthreshold region of the device, where the device operates in a weakly conducting state. It can be judged from the turning point of drain curve ($I_d$) that the cut-off voltage ($V_{th}$) is about -9.5 V. Combining with the ferromagnetism above 300 K, this **U1@U2** film belongs definitely to an excellent intrinsic room-temperature magnetic semiconductor.

The AHE phenomena of the **U1@U2** film at zero magnetic field is also tested (Fig. 6d). The Hall voltage ($V_H$) can be obtained by applying different currents in the longitudinal direction under zero magnetic field. It can be found that the polarity of $V_H$ reverses with the switching of the current direction, while its magnitude remains unchanged. This feature helps eliminate totally the lateral voltage deviations caused by electrode asymmetry [45]. The anomalous Hall resistivity ($\rho_H$) of the **U1@U2** film is measured to be 0.32 mΩ·cm at room-temperature, which is 1~2 orders of magnitude higher than that of recently reported magnetic materials [46–48]. The remarkable AHE at room temperature implies that the crystal has strong spin-orbit coupling and magnetically ordered structures [49], laying the foundation for ferrotoroidicity.

## III. Conclusion

This work is distinguished by several key advances. First, the preparation of room-temperature ferrotoroidic material is realized for the first time by supramolecular self-assembly method. The material presents the main characteristics required by the ferrotoroidicity. The mechanism from microscopic to macroscopic levels is revealed. Secondly, the material exhibits simultaneously excellent magnetic and semiconductor properties at an ambient condition, which belongs to a unique room-temperature magnetic semiconductor designed by an effective approach avoiding ordinary doping. Thirdly, the ferrotoroidic crystalline film fabricated presents superhigh hole mobility and remarkable room-temperature AHE phenomenon. A successful example of ferrotoroidicity, the fourth primary ferroic order, plays an important role in promoting the development of condensed matter physics. The uranyl based ferrotoroidicity helps understand the basic properties of actinides related to 5f electron and the spin in terms of a new perspective. As far as the applications, this high-quality ferrotoroidic film collecting several specific properties such as the marked room-temperature AHE and superhigh hole mobility, is of essential in exploring spin-orbit torque MRAM, especially as a flexible material.

This work is supported by Beijing National Laboratory for Molecular Science. Authors are grateful to Professor Yanfeng Zhang's group for the help in SHG measurements. Authors thank the molecular materials and nanomachining laboratory of Peking University. The high-performance computing platform at Peking University is acknowledged.

*Corresponding author: xshen@pku.edu.cn.